\documentclass[journal,draftcls,onecolumn,12pt,english]{IEEEtran}

\usepackage{amsthm}
\usepackage{amsmath}
\usepackage{amssymb}
\usepackage{multirow}
\usepackage{mathrsfs}
\usepackage{graphicx}
\usepackage{slashbox}
\usepackage[T1]{fontenc}
\usepackage[utf8]{inputenc}

\theoremstyle{plain}
\newtheorem{thm}{\protect\theoremname}
\theoremstyle{plain}
\newtheorem{lem}[thm]{\protect\lemmaname}
\theoremstyle{plain}
\newtheorem{cor}[thm]{\protect\corollaryname}

\usepackage{babel}
\providecommand{\lemmaname}{Lemma}
\providecommand{\theoremname}{Theorem}
\providecommand{\corollaryname}{Corollary}

\newcommand{\cB}{\mathcal{B}}
\newcommand{\cI}{\mathcal{I}}
\newcommand{\cC}{\mathcal{C}}

\begin{document}

\title{On the Optimal Minimum Distance of \\ Fractional Repetition Codes}

\author{Bing~Zhu,~\IEEEmembership{Member,~IEEE,}
        Kenneth~W.~Shum,~\IEEEmembership{Senior Member,~IEEE,}
        Weiping~Wang,~\IEEEmembership{Member,~IEEE,}
        and~Jianxin~Wang,~\IEEEmembership{Senior Member,~IEEE}

\thanks{B. Zhu, W. Wang, and J. Wang are with the School of Computer Science and Engineering, Central South University, Changsha 410083, China (e-mail: zhubing@csu.edu.cn; wpwang@csu.edu.cn; jxwang@csu.edu.cn).}
\thanks{K. W. Shum is with the School of Science and Engineering, Chinese University of Hong Kong (Shenzhen), Shenzhen 518172, China (e-mail: wkshum@inc.cuhk.edu.hk).}}

\maketitle

\begin{abstract}
Fractional repetition (FR) codes are a class of repair efficient erasure codes that can recover a failed storage node with both optimal repair bandwidth and complexity. In this paper, we study the minimum distance of FR codes, which is the smallest number of nodes whose failure leads to the unrecoverable loss of the stored file. We consider upper bounds on the minimum distance and present several families of explicit FR codes attaining these bounds. The optimal constructions are derived from regular graphs and combinatorial designs, respectively.
\end{abstract}

\begin{IEEEkeywords}
Distributed storage, fractional repetition codes, minimum distance, regular graphs, combinatorial designs.
\end{IEEEkeywords}

\section{Introduction}

Modern distributed cloud storage systems that built on large number of independent storage devices can provide large-scale storage services in a cost-efficient manner. However, due to~the commodity nature of physical storage nodes, component failures may occur unexpectedly (and even frequently) in such systems. To ensure data availability, distributed storage systems need to introduce a certain level of data redundancy in order to protect the stored data against node failures. A simple solution that is deployed in realistic systems is to store several replicas of each data object, and upon failure of a single storage node, the lost data can be exactly recovered by downloading the remaining replicas from other nodes. Data replication supports efficient~repair and is easy to manage in practice, yet it suffers from the main drawback of high storage overhead. Instead, erasure coding has emerged as a promising technology for distributed storage systems with the advantage of achieving higher storage efficiency~\cite{key-1}. For example, an $[n, k]$ maximum distance separable (MDS) code encodes a data file consisting of $k$ symbols into $n$ coded blocks in such a manner that the original data can be reconstructed by accessing any $k$ out of the $n$ coded blocks (this property is called the \textit{MDS property}). In this traditional coding framework, a failed node can be repaired by contacting $k$ surviving nodes and then re-encoding the lost data. This recovery method turns out to~be bandwidth consuming since the repair of one block incurs the transfer of $k$ coded blocks over the storage network.

Regenerating codes are introduced in~\cite{key-2} with the capability of minimizing the network traffic required for repairing a failed storage node. In an $ [n,k,d,\alpha,\beta]$ regenerating code, a data object is encoded into $n\alpha$ coded packets, which are distributed across $n$ storage nodes, each containing $\alpha$ packets. The source data can be reconstructed by contacting any $k$ nodes in the system, where $k$ is called the \textit{reconstruction degree}. When a node fails, the lost packets can be regenerated by connecting to any subset of $d$ surviving nodes and downloading $\beta$ packets from each node. The number of nodes contacted for repair (i.e., $d$) is called the \textit{repair locality}, and the total number of packet transmissions (i.e., $d\beta$) is called the \textit{repair bandwidth}. If the repair bandwidth equals to the storage capacity $\alpha$ of the failed node, the corresponding code~is called a \textit{minimum-bandwidth regenerating (MBR)} code. In the regenerating framework proposed in~\cite{key-2}, each set of $\beta$ packets is obtained as linear combinations of the $\alpha$ packets stored in the connected node, which increases the computational complexity~of repair. For the scenario that the $\beta$ transferred packets are taken as subsets of the $\alpha$ packets in the helper nodes, the repair regime is called \textit{repair-by-transfer}~\cite{key-3} or \textit{uncoded repair}~\cite{key-4}. In other words, this repair approach enjoys the same efficiency as data replication. Erasure codes with this desirable property can be found in~\cite{key-3}\textendash{}\cite{key-10}.

Fractional repetition (FR) codes are a special class of MBR codes proposed in~\cite{key-4} that enable uncoded repairs of failed storage nodes. It has a tailor-made encoding architecture in which an outer MDS code first performs the encoding operation of data objects and then an inner FR code replicates and distributes the coded packets across the storage nodes in a sophisticated manner. The stored file can be recovered by downloading a sufficient number of distinct coded packets and decoding the original data according to the MDS property. Moreover, the repair of a failed storage node can be completed by contacting some specific sets of helper nodes that contain the remaining replicas of lost packets, which differs from conventional MBR codes wherein any set of $d$ surviving nodes are eligible for node repair.

Consider a distributed storage system consisting of $n$ nodes, where each node stores the same number of $\alpha$ packets. Suppose that the size of stored data objects is $M$. The \textit{minimum distance} of this storage system, denoted by $d_{\min}$, is the number of nodes such that
\begin{enumerate}
\item there exists at least one set of $d_{\min}$ nodes whose erasure leads to the unrecoverable loss~of the source data;
\item the stored file can be recovered for any subset of $d_{\min} - 1$ node erasures.
\end{enumerate}

In~\cite{key-11}, Kamath~\textit{et al.} derived a Singleton-like bound on~the minimum distance $d_{\min}$ as follows
\begin{equation} \label{SingletonBound}
d_{\min} \leq n - \Big\lceil \frac{M}{\alpha} \Big\rceil + 1.
\end{equation}
Furthermore, Papailiopoulos and Dimakis showed in~\cite{key-12} that erasure codes with the local repair property (i.e., $d<k$) will incur a penalty on the maximum possible minimum distance. They proved that for an erasure code with repair locality $d$, the minimum distance of the corresponding storage system is upper bounded by
\begin{equation} \label{Papailiopoulos}
d_{\min} \leq n - \Big\lceil \frac{M}{\alpha} \Big\rceil - \Big\lceil \frac{M}{d\alpha} \Big\rceil + 2.
\end{equation}
Clearly, the upper bound in~\eqref{Papailiopoulos} reduces to that in~\eqref{SingletonBound} when $k \leq d$. To the best of our knowledge, existing FR codes attaining the upper bounds above are known for some scattered examples~\cite{key-13} or small reconstruction degrees (e.g. $1\leq k \leq 3$)~\cite{key-14,key-15}.

\begin{table*}
\protect\caption{Summary of Optimal FR Codes Attaining the Singleton-like Bound in~\eqref{SingletonBound}, Where $p$ Is a Prime and $q$ Is a Prime Power}
\begin{centering}
\begin{tabular}{|c|c|c|c|}
\hline
Method & Code parameter & File size & Requirements \tabularnewline
\hline
\hline
Theorem~\ref{thm:regular} & $(n,\alpha,2)$  & $k\alpha-k+1$ & $1\leq k \leq \min\{\alpha, g-1\}$ \tabularnewline
\hline
Corollary~\ref{cor:regular} & $(n,\alpha,2)$ & $k\alpha-k$ & $g\leq k \leq \min\{\alpha-1, g+\lceil\frac{g}{2}\rceil-2\}$ \tabularnewline
\hline
Theorem~\ref{thm:Turán} & $(n,\frac{n(r-1)}{r},2)$ & $\frac{k(r-1)n}{r} - \Big\lfloor \frac{r-1}{r} \cdot \frac{k^2}{2} \Big\rfloor$ & $1\leq k \leq \min\{\frac{n(r-1)}{r}, \sqrt{2n} - 1\}$ \tabularnewline
\hline
Theorem~\ref{thm:Steiner} & $(\theta, \rho = \frac{\theta-1}{\alpha-1},\alpha)$ & $k\rho - \binom{k}{2}$ & $1\leq k \leq \lceil \frac{\sqrt{1+8\rho} - 1}{2} \rceil$ \tabularnewline
\hline
\multirow{2}{*}{Theorem~\ref{thm:affine}} & \multirow{2}{*}{$(q\rho, q^{m-1}, \rho \leq \frac{q^m-1}{q-1})$}  & \multirow{2}{*}{ $q^m \big[ 1-(1-\frac{1}{q})^k \big]$ } & $1\leq k \leq \min \{m, k_0\}$, where $k_0$ is the largest integer \tabularnewline
 &  &  &  such that Inequality~\eqref{condition_affine} holds. \tabularnewline
\hline
Corollary~\ref{cor:MOLS} & $(\rho p^m, p^m, \rho \leq p^m-1)$ & $kp^m - \binom{k}{2}$ & $1\leq k \leq \min\{\rho, \lceil \frac{\sqrt{1+8p^m} - 1}{2} \rceil\}$ \tabularnewline
\hline
\end{tabular}
\par\end{centering}
\label{Summary:OFRC1}
\end{table*}

\textit{Contributions:} In this paper, we study upper bounds on the minimum distance of FR codes and present explicit code constructions that attain these bounds. Specifically, the main contributions are summarized as follows.

\begin{itemize}
  \item We show that the Singleton-like upper bound in~\eqref{SingletonBound} is equivalent to a simple lower bound on the reconstruction degree of FR codes. Based on this relation, we present several families~of optimal FR codes attaining the bound in~\eqref{SingletonBound}, which are summarized in Table~\ref{Summary:OFRC1}. For the given file size, the minimum distance of the corresponding system~is optimal with respect to the Singleton-like bound in~\eqref{SingletonBound}.%
      \footnote{In~\cite{key-13,key-14}, the authors showed via examples that the constructed FR codes attain the bound in~\eqref{SingletonBound}, where the corresponding reconstruction degrees are mainly within the range $1\leq k \leq 3$. As summarized in Table~\ref{Summary:OFRC1}, our proposed constructions support a wide range of code parameters.}
  \item We propose explicit constructions of FR codes attaining the minimum distance bound in~\eqref{Papailiopoulos}, i.e, optimal FR codes that also have the local repair property. The proposed constructions are derived from regular graphs with large girth.
  \item We derive an improved upper bound on the minimum distance of FR codes, which is tighter than the existing bounds especially for some large file sizes. Moreover, we also discuss some optimal FR codes attaining the proposed upper bound.
\end{itemize}

\smallskip{}

\textit{Organization:} The remainder of this paper is organized as follows. Section II provides a brief overview of FR codes, including the necessary definitions, properties, and related works. Section III introduces several families of FR codes that achieve the Singleton-like bound in~\eqref{SingletonBound}. Section IV proposes optimal FR codes attaining the upper bound in~\eqref{Papailiopoulos}. Section V presents a new upper bound on the minimum distance of FR codes and compares this bound to the existing bounds. Finally, Section VI concludes the paper.

\section{Preliminaries}

\subsection{Fractional Repetition Codes}

An \textit{incidence structure} is a triple $(X,\cB,\cI)$, where $X$~and~$\cB$ are nonempty (finite) sets and $\cI$ is a subset of $X\times\cB$, i.e., $\cI \subseteq X\times\cB$. The elements in~$X$ are called \textit{points}, and the elements in $\cB$ are called \textit{blocks}. We say that a point $x\in X$ is \textit{incident} with a block $B\in \cB$ if $(x,B)\in\cI$. An \textit{$(n,\alpha,\rho)$-FR code} $\cC$ is an incidence structure $(X,\cB,\cI)$ with $|\cB|=n$ such that each point is incident with $\rho$ blocks in $\cB$ and each block is incident with $\alpha$ points in $X$. Thus, the number of points in $X$ is $\theta:=\frac{n\alpha}{\rho}$. Moreover, the \textit{dual code} of~$\cC$, denoted by $\cC^t$, is the incidence structure $(\cB,X,\cI^t)$, where $\cI^t$ is the subset of $\cB\times X$ defined by
$$
\cI^t := \{(B,x):\, (x,B)\in \cI\}.
$$
Since the roles of points and blocks are reversed in $\cI^t$, it follows that $\cC^t$ forms a $(\theta,\rho,\alpha)$-FR code~\cite{key-16}.

To store a data object of size $M$, we first perform the encoding operation by adopting a $[\theta, M]$ MDS code and then spread the coded packets across $n$ storage nodes using an $(n,\alpha,\rho)$-FR code $\cC=(X,\cB,\cI)$ in the following manner: each coded packet is associated with a point in $X$ and each storage node is associated with a block in $\cB$. Therefore, each packet is equally replicated $\rho$ times and each node contains exactly $\alpha$ packets. According to the MDS property, we need to collect $M$ distinct coded packets for data retrieval, and the smallest number of nodes in $\cC$ that cover at least $M$ distinct packets gives the \textit{reconstruction degree} of $\cC$. In this sense, the value of $M$ (i.e., the dimension of the outer MDS code), is closely related to the reconstruction degree of the inner FR code, which motivates the following definition.

For a given reconstruction degree $k$, the \textit{supported file size} of $\cC=(X,\cB,\cI)$, denoted by $M_k(\cC)$, is defined as
$$
M_k(\cC) := \min_{\mathcal{K}\subset \cB, |\mathcal{K}| = k} |\{x\in X:\, \exists B \in \mathcal{K}, (x, B) \in \cI \}|.
$$
Intuitively, the value of $M_k(\cC)$ refers to the smallest number~of distinct points in any subset of $k$ blocks in $\cB$. Thus by setting $M_k(\cC)$ as the dimension of the outer MDS code, we can decode the stored data by contacting arbitrary $k$ storage nodes in $\cC$. For $M \leq \theta$, let $k$ be the reconstruction degree such that $M_k(\cC) \geq M$. Then, the minimum distance of $\cC$-based system is $n-k+1$ when the size of stored data object is $M$.

However, it is a non-trivial task to calculate the file size of FR codes with given parameters. In~\cite{key-4}, a tight upper bound on the supported file size of an $(n,\alpha,\rho)$-FR code $\cC$ is given as
\begin{equation} \label{eq:FRbound}
M_k(\cC) \leq \varphi(k),
\end{equation}
where $\varphi(k)$ is defined recursively by
\begin{equation} \label{eq:FRbound2}
\varphi(1) := \alpha, \ \varphi(k+1) := \varphi(k) + \alpha - \Big\lceil \frac{\rho \varphi(k) - k \alpha}{n-k}\Big\rceil.
\end{equation}

For example, consider now the $(7,3,3)$-FR code $\cC$ with $X=\{1,2,\ldots,7\}$, and $\cB= \{\{1,2,4\},$ $\{1,3,7\},\{1,5,6\},\{2,3,5\},\{2,6,7\},\{3,4,6\},\{4,5,7\}\}$. It can be computed that
$$
M_k(\cC) = \begin{cases}
7, & \text{for } k= 5,6,7, \\
6, & \text{for } k = 3, 4,\\
5, & \text{for } k = 2, \\
3, & \text{for } k = 1.
\end{cases}
$$
We note here that $\cC$ achieves the upper bound on the file size in~\eqref{eq:FRbound} for $1\leq k \leq 7$, yet it attains the Singleton-like bounds in~\eqref{SingletonBound} and~\eqref{Papailiopoulos} only for $1\leq k \leq 2$.

In~\cite{key-16}, the \textit{complementary supported file size} of an $(n,\alpha,\rho)$-FR code $\cC$, denoted by $N_k(\cC)$,~is defined as follows
$$
N_k(\cC) := \frac{n\alpha}{\rho} - M_k(\cC).
$$
In contrast, $N_k(\cC)$ refers to the size of the largest set of packets that are not covered by $k$ storage nodes in $\cC$. By virtue of $N_k(\cC)$, the elementary relationship between the supported file size of an FR code and its dual is revealed, as detailed in the following lemma.

\begin{lem} (\cite{key-16}) \label{dualrelation}
Let $\cC$ be an $(n,\alpha,\rho)$-FR code with $\theta= \frac{n\alpha}{\rho}$, and let $\cC^t$ be the dual code. Then, the supported file size of $\cC$ can be determined as
\begin{equation}
M_k(\cC) = \begin{cases}
\theta-1, & \text{for } N_2(\cC^t) <k \leq  N_1(\cC^t),\\
\theta-2, & \text{for } N_3(\cC^t) <k \leq  N_2(\cC^t),\\
\vdots & \vdots \\
2, & \text{for } N_{\theta-1}(\cC^t) <k \leq  N_{\theta-2}(\cC^t),\\
1, & \text{for } N_\theta(\cC^t)=0 <k \leq  N_{\theta-1}(\cC^t).
\end{cases}
\label{eq:Mk}
\end{equation}
\end{lem}

\subsection{Related Work}

The construction of FR codes has attracted considerable attention over the past decade, which is mainly derived from regular graphs and combinatorial designs. In the pioneer work in~\cite{key-4}, the authors provided explicit constructions of FR codes based on regular graphs and Steiner~systems. The constructions from bipartite cage graphs in~\cite{key-17} have the advantage that the corresponding system can be easily expanded without frequent reconfigurations. In~\cite{key-18}, Silberstein and Etzion presented some optimal FR codes that achieve the upper bound on the supported file size in~\eqref{eq:FRbound}, which are constructed from incidence structures including extremal graphs, transversal designs and generalized polygons. Constructions of FR codes from partially ordered sets are considered in~\cite{key-19}, wherein the resulting codes can store larger files than MBR codes for the same system parameters. In~\cite{key-13}, Olmez and Ramamoorthy investigated constructions of FR codes based on resolvable combinatorial designs including affine geometries, Hadamard designs, and mutually orthogonal Latin squares. Constructions of new FR codes from existing codes can be obtained by using techniques such as Kronecker product~\cite{key-13}, tensor product~\cite{key-16}, and symbol extension~\cite{key-20}. FR codes that support local repair are studied in~\cite{key-15} and~\cite{key-21}, which are devised from bipartite graphs and symmetric designs, respectively. Moreover, the FR codes proposed~in~\cite{key-22} enjoy the additional property that each helper node contributes the same amount of data during the repair of multiple failed nodes, and FR codes attaining lower bounds on the reconstruction degree are discussed in~\cite{key-23}.

Despite the rich source for code constructions, it remains a challenging task to calculate the supported file size of FR codes, which is essentially equivalent to the problem of determining the expansion of bipartite graphs~\cite{key-13}. Among the constructions above, only a few have determined the supported file size of designed FR codes for certain parameter ranges due to the algebraic properties of corresponding constructions.

\section{Optimal FR Codes Attaining the Singleton-like Bound in~\eqref{SingletonBound}}

In this section, we consider explicit constructions of FR codes that achieve the Singleton-like bound in~\eqref{SingletonBound}. We begin with the following useful theorem that establishes a connection between the Singleton-like bound on the minimum distance $d_{\min}$ and a lower bound on the reconstruction degree $k$ of FR codes.

\begin{thm} \label{basecondition}
Let $\cC$ be an $(n,\alpha,\rho)$-FR code. Then, we have
\begin{equation} \label{bound_k}
k \geq \Big\lceil \frac{M_k(\cC)}{\alpha} \Big\rceil,
\end{equation}
with equality holds if and only if
$d_{\min} = n - \lceil \frac{M_k(\cC)}{\alpha} \rceil + 1$.
\end{thm}

\begin{IEEEproof}
Since each storage node in $\cC$ contains $\alpha$ coded packets,~we obtain $k\alpha$ packets from any collection of $k$ nodes, which gives that $M_k(\cC) \leq k \alpha$. As $k$ is an integer, the inequality in~\eqref{bound_k} follows. Because the stored file can be recovered from any $k$ nodes, thus the system can tolerate $n-k$ node failures, i.e., $d_{\min} - 1 \leq n - k$.

Suppose $k = \lceil \frac{M_k(\cC)}{\alpha} \rceil$. We can write $M_k(\cC) = k \alpha - t$ for some integer $0 \leq t < \alpha$. In this~case, it follows that $(k-1)\alpha < M_k(\cC)$, meaning that any $k-1$ nodes do not contain enough packets for successful decoding. This implies $d_{\min} = n-k+1 = n - \lceil \frac{M_k(\cC)}{\alpha} \rceil + 1$.

Conversely, suppose now $d_{\min} = n - \lceil \frac{M_k(\cC)}{\alpha} \rceil + 1$. Because $d_{\min} \leq n - k  + 1$ holds in general, we obtain $n - \lceil \frac{M_k(\cC)}{\alpha} \rceil + 1 \leq   n - k  + 1$, which is equivalent to $k \leq \lceil \frac{M_k(\cC)}{\alpha} \rceil$.
\end{IEEEproof}

Based on the result above, our main objective in this section is to design FR codes such that the following relation holds
\begin{equation} \label{Condition}
k = \Big\lceil \frac{M_k(\cC)}{\alpha} \Big\rceil.
\end{equation}
Clearly, the following result holds for $k=1,2$.
\begin{lem}
An FR code $\cC$ attains the Singleton-like bound in~\eqref{SingletonBound} for $k=1,2$ if there are no repeated blocks in $\cC$.
\end{lem}

In the following discussions, we mainly focus on optimal FR codes with reconstruction degree $k\geq 3$. Since an incidence structure is a basic concept in the theory of graphs and combinatorial designs, we discuss code constructions from the two methods separately.

\subsection{Optimal Constructions from Regular Graphs}

A graph $G$ consists of a vertex set and an edge set, and two vertices are said to be \textit{incident} if there exists an edge between them. The \textit{degree} of a vertex is defined as the number of edges that are incident with it, and if every vertex in $G$ has the same degree of $\alpha$, then $G$ is called an \textit{$\alpha$-regular graph}. In particular, Turán graphs are a special family of regular graphs defined as follows. Let $n$ and $r$ be two integers such that $r$ divides $n$.~An \textit{$(n, r)$-Turán graph} is formed by partitioning a set~of $n$ vertices into $r$ distinct subsets, and connecting two vertices by an edge if they belong to different subsets. It follows that the degree~of each vertex is $(r-1)\frac{n}{r}$.

In graph theory, a \textit{cycle} is a nonempty trail in which the~only repeated vertices are the first and last vertices (i.e., each vertex has degree two). The \textit{length} of a cycle is its number of edges, and the \textit{girth} of a graph is the minimum length of a cycle in~the graph. As a concrete example, the girth of an $(n, r)$-Turán graph with $n\geq 4$ and $r=2$ is $4$, and if $r\geq 3$, the girth is $3$.

The construction rationale of regular graph based method is to treat each vertex as a block and each edge as a point.~Hence, each packet in the resulting FR code is replicated twice and~the degree of each vertex determines the capacity of storage nodes, i.e., an $\alpha$-regular graph with $n$ vertices can be adopted to yield an $(n,\alpha,2)$-FR code. In~\cite{key-18}, Silberstein and Etzion derived the supported file~size of the FR codes based on Turán graphs and regular graphs with large girth. Specifically, the file size of the FR code $\cC$ constructed from an $\alpha$-regular graph with girth $g$ is
\begin{equation}
M_k(\cC) = \begin{cases}
k\alpha-k+1, & \text{for } 1 \leq k\leq g-1 \\
k\alpha-k, & \text{for } g \leq k \leq g+\lceil\frac{g}{2}\rceil-2
\end{cases}
\label{FileSizeRG}
\end{equation}
and an $(n, r)$-Turán graph based FR code $\cC$ has file size
\begin{equation}
M_k(\cC) = \frac{k(r-1)n}{r} - \Big\lfloor \frac{r-1}{r} \cdot \frac{k^2}{2} \Big\rfloor,
\end{equation}
for $1 \leq k \leq (r-1)\frac{n}{r}$.

\begin{thm} \label{thm:regular}
Let $G$ be an $\alpha$-regular graph with girth $g$. Then, the FR code $\cC$ based on $G$ attains the Singleton-like bound in~\eqref{SingletonBound} for $1\leq k \leq \min\{\alpha, g-1\}$.
\end{thm}

\begin{IEEEproof}
Since the file size of $\cC$ is $M_k(\cC) = k\alpha-(k-1)$~for $1 \leq k\leq g-1$, we have
\begin{equation} \label{girth}
\Big\lceil \frac{M_k(\cC)}{\alpha} \Big\rceil = k - \Big\lfloor \frac{k-1}{\alpha} \Big\rfloor.
\end{equation}
The right-hand side term in~\eqref{girth} equals to $k$ if
\begin{equation}
0 \leq \frac{k-1}{\alpha} < 1,
\end{equation}
which gives that
\begin{equation}
1 \leq k < \alpha+1.
\end{equation}

This completes the proof.
\end{IEEEproof}

\smallskip{}

\textit{Remark 1.} It is shown in~\cite{key-18} that the FR code $\cC$ based~on an $\alpha$-regular graph with girth $g \geq \alpha + 1$ is optimal with respect to the upper bound on the supported file size in~\eqref{eq:FRbound}. We note that in this case, $\cC$ is also optimal with respect to the Singleton-like bound in~\eqref{SingletonBound} for $1\leq k \leq \alpha$.

The following corollary is an immediate consequence from the discussion above.

\begin{cor} \label{cor:regular}
Suppose that $G$ is an $\alpha$-regular graph with girth $g <\alpha$. Then, the FR code based on $G$ attains the Singleton-like bound in~\eqref{SingletonBound} for $g\leq k \leq \min\{\alpha-1, g+\lceil\frac{g}{2}\rceil-2\}$.
\end{cor}

We illustrate the power of the results above by evaluating~FR codes from the incidence graph of projective planes. Let $q\geq 2$ be a prime power. The incidence graph of a projective plane~of order $q$ is a $(q+1)$-regular graph that consists of $2q^2+2q+2$ vertices~\cite{key-24}. Furthermore, the girth of this incidence graph is $g=6$. Then,
\begin{enumerate}
\item if $2\leq q \leq 5$, there exists a $(2q^2+2q+2,q+1,2)$-FR code attaining the Singleton-like bound in~\eqref{SingletonBound} for $1\leq k \leq \min\{5, q+1\}$;
\item if $q \geq 7$, there exists a $(2q^2+2q+2,q+1,2)$-FR code attaining the Singleton-like bound in~\eqref{SingletonBound} for $6\leq k \leq 7$.
\end{enumerate}

\smallskip{}

Next, we proceed to consider constructions of FR codes derived from Turán graphs. From a practical perspective, we focus on $(n, r)$-Turán graphs with $n\geq 4$ and $r\geq 2$.

\begin{thm} \label{thm:Turán}
An $(n, r)$-Turán graph based FR code attains the Singleton-like bound in~\eqref{SingletonBound} for $1\leq k \leq \min\{\frac{n(r-1)}{r}, \sqrt{2n} - 1\}$.
\end{thm}

\begin{IEEEproof}
Note that each node contains $\alpha=(r-1)\frac{n}{r}$ packets, thus we need to prove that
\begin{equation} \label{Turán}
\Big\lfloor \frac{r-1}{r} \cdot \frac{k^2}{2} \Big\rfloor <\frac{n(r-1)}{r}.
\end{equation}

Since the right-hand side term in the inequality is a positive integer, we can remove the floor operator, i.e.,
\begin{equation}
k^2 < 2n,
\end{equation}
which completes that proof.
\end{IEEEproof}

\smallskip{}

For example, a $(50, 5)$-Turán graph based FR code attains the Singleton-like bound in~\eqref{SingletonBound} for $1\leq k \leq 9$.

\subsection{Optimal Constructions from Combinatorial Designs}

A \textit{combinatorial design} (or \textit{design}) is an incidence structure $(X,\cB,\cI)$, in which the blocks in $\cB$ are a collection~of subsets of $X$ whose intersections have specified numerical properties. In design theory, different types of combinatorial designs have been introduced with the block intersection numbers satisfying certain requirements~\cite{key-25}. For example, a \textit{Steiner system $S(2,\alpha,\theta)$} is a set $X$ of $\theta$ points together with a family $\cB$ of $\alpha$-subsets of $X$ with the property that every pair of points in $X$ is contained in exactly one block. By a simple counting argument, it follows that each point is incident with $\rho=\frac{\theta-1}{\alpha-1}$ blocks. Moreover, the largest subset of $X$ which intersects every block in $\cB$ in~either zero or two points is~called the \textit{maximal arc} in $S(2,\alpha,\theta)$.

A design $(X,\cB,\cI)$ is said to be \textit{resolvable} if its block set~$\cB$ can be partitioned into several parallel classes, each of which is a set of blocks that partition the point set $X$. If any two blocks from different parallel classes intersect in a constant number~of points, then such a design is called an \textit{affine resolvable design}~\cite{key-25}.

By comparing the definitions of FR codes and combinatorial designs, we observe that any design satisfying the property that each block contains the same number of points and each point occurs in the same number of blocks can be leveraged to yield an FR code.

We start by investigating FR codes constructed from Steiner systems. Note that the dual of an $S(2,\alpha,\theta)$ based FR code $\cC$~is a $(\theta,\rho,\alpha)$-FR code with $\rho=\frac{\theta-1}{\alpha-1}$. If the applied Steiner system has a maximal arc of size $\rho+1$, then the supported file size~of $\cC^t$ is shown in~\cite{key-13} to be
\begin{equation} \label{FileSizeSteiner}
M_k(\cC^t) = k\rho-\binom{k}{2},
\end{equation}
where $1 \leq k \leq \rho + 1$.

\begin{thm} \label{thm:Steiner}
Let $\cC$ be an FR code constructed from a Steiner system $S(2,\alpha,\theta)$ with $\rho=\frac{\theta-1}{\alpha-1}$, such that it has a maximal~arc of size $\rho+1$. Then, $\cC^t$ achieves the Singleton-like bound in~\eqref{SingletonBound} for $1\leq k \leq \lceil \frac{\sqrt{1+8\rho} - 1}{2} \rceil$.
\end{thm}

\begin{IEEEproof}
Considering the file size of $\cC^t$ in~\eqref{FileSizeSteiner}, it remains~to prove that
\begin{equation}
\rho > \binom{k}{2},
\end{equation}
since each node in $\cC^t$ contains $\rho$ packets. Therefore, we obtain
\begin{equation} \label{intermediate_Steiner}
1\leq k < \frac{1+\sqrt{1+8\rho}}{2}.
\end{equation}
The proof is completed since the right-hand side term in~\eqref{intermediate_Steiner}~is strictly smaller than $\rho+1$.
\end{IEEEproof}

\begin{table}
\caption{Optimal FR Codes Derived from Steiner Systems}
\begin{centering}
\begin{tabular}{c|c|c|c}
\hline
$\alpha$ & Steiner system & Code parameter & Range of $k$\tabularnewline
\hline
\hline
\multirow{9}{*}{$3$} & $S(2,3,15)$ & $(15,7,3)$ & $1\leq k\leq4$\tabularnewline
\cline{2-4} \cline{3-4} \cline{4-4}
 & $S(2,3,19)$ & $(19,9,3)$ & $1\leq k\leq4$\tabularnewline
\cline{2-4} \cline{3-4} \cline{4-4}
 & $S(2,3,27)$ & $(27,13,3)$ & $1\leq k\leq5$\tabularnewline
\cline{2-4} \cline{3-4} \cline{4-4}
 & $S(2,3,31)$ & $(31,15,3)$ & $1\leq k\leq5$\tabularnewline
\cline{2-4} \cline{3-4} \cline{4-4}
 & $S(2,3,39)$ & $(39,19,3)$ & $1\leq k\leq6$\tabularnewline
\cline{2-4} \cline{3-4} \cline{4-4}
 & $S(2,3,43)$ & $(43,21,3)$ & $1\leq k\leq6$\tabularnewline
 \cline{2-4} \cline{3-4} \cline{4-4}
 & $S(2,3,51)$ & $(51,25,3)$ & $1\leq k\leq7$\tabularnewline
 \cline{2-4} \cline{3-4} \cline{4-4}
 & $S(2,3,55)$ & $(55,27,3)$ & $1\leq k\leq7$\tabularnewline
 \cline{2-4} \cline{3-4} \cline{4-4}
 & $S(2,3,63)$ & $(63,31,3)$ & $1\leq k\leq8$\tabularnewline
\hline
\multirow{9}{*}{$4$} & $S(2,4,13)$ & $(13,4,4)$ & $1\leq k\leq3$\tabularnewline
\cline{2-4} \cline{3-4} \cline{4-4}
 & $S(2,4,16)$ & $(16,5,4)$ & $1\leq k\leq3$\tabularnewline
\cline{2-4} \cline{3-4} \cline{4-4}
 & $S(2,4,25)$ & $(25,8,4)$ & $1\leq k\leq4$\tabularnewline
\cline{2-4} \cline{3-4} \cline{4-4}
 & $S(2,4,28)$ & $(28,9,4)$ & $1\leq k\leq4$\tabularnewline
\cline{2-4} \cline{3-4} \cline{4-4}
 & $S(2,4,40)$ & $(40,13,4)$ & $1\leq k\leq5$\tabularnewline
\cline{2-4} \cline{3-4} \cline{4-4}
 & $S(2,4,49)$ & $(49,16,4)$ & $1\leq k\leq6$\tabularnewline
\cline{2-4} \cline{3-4} \cline{4-4}
 & $S(2,4,52)$ & $(52,17,4)$ & $1\leq k\leq6$\tabularnewline
\cline{2-4} \cline{3-4} \cline{4-4}
 & $S(2,4,76)$ & $(76,25,4)$ & $1\leq k\leq7$\tabularnewline
\cline{2-4} \cline{3-4} \cline{4-4}
 & $S(2,4,88)$ & $(88,29,4)$ & $1\leq k\leq8$\tabularnewline
\hline
\end{tabular}
\par\end{centering}
\label{OFRSS}
\end{table}

\textit{Remark 2.} Except for finitely~many constructions of $S(2,\alpha,\theta)$ with $\alpha > 5$, Steiner systems are mainly known~to exist for small values of $\alpha$, i.e., $\alpha= 2,3,\ldots,5$. Furthermore, there are no many general results regarding the existence of maximal arcs in Steiner systems. However, it is worth noting here that a Steiner system $S(2,3,\theta)$ has at least one maximal arc if $\theta \geq 7$ and $\theta \equiv 3,7 \textup{ (mod 12)}$~\cite{key-26}, and a Steiner system $S(2,4,\theta)$ has a maximal arc of size $\frac{\theta+2}{3}$ if $\theta \geq 13$, $\theta \equiv 1,4 \textup{ (mod 12)}$, and $\frac{\theta-1}{3}$ is a prime power~\cite{key-27}. Indeed, these two infinite families of Steiner systems that have maximal arcs can be of practical interest for real-world systems since the duals of designed FR codes have an applicable repetition degree of $3$ or $4$.

Table~\ref{OFRSS} lists several explicit optimal FR codes derived from Steiner systems $S(2,3,\theta)$ and $S(2,4,\theta)$ respectively. In addition to the results above, we can also construct FR codes that are optimal with respect to the Singleton-like bound in~\eqref{SingletonBound} for relatively large values of $k$, e.g., there exists a $(387,193,3)$-FR code achieving the Singleton-like bound in~\eqref{SingletonBound} for $1\leq k \leq 20$.

\smallskip{}

In what follows, we discuss another construction of FR codes from affine resolvable designs. Let $q$ be a prime power and $m \geq 2$ be an integer. Based on affine geometries, a family of affine resolvable designs is introduced in~\cite{key-13}, which contains $\frac{q^m-1}{q-1}$ parallel classes. Using this design, the authors devised a $(q\rho, q^{m-1}, \rho \leq \frac{q^m-1}{q-1})$-FR code $\cC$ whose file size is
\begin{equation}
\label{filesize:affine}
M_k(\cC) = q^m \big[ 1-(1-\frac{1}{q})^k \big],
\end{equation}
where $1 \leq k \leq m$. In the derivation of $M_k(\cC)$, the parameter $\rho$ should be chosen in such a manner that if $q>m$, then $\rho> m$ and if $q \leq m$, then $\rho \leq m$.

\begin{thm} \label{thm:affine}
Let $q$ be a prime power and $m \geq 2$ be an integer. Let $k_0$ denote the largest integer such that
\begin{equation}
\label{condition_affine}
q(1-\frac{1}{q})^{k_0} + k_0 - q < 1.
\end{equation}
Then, there exists a $(q\rho, q^{m-1}, \rho \leq \frac{q^m-1}{q-1})$-FR code that attains the Singleton-like bound in~\eqref{SingletonBound} for $1\leq k \leq \min \{m, k_0\}$.
\end{thm}

\begin{IEEEproof}
By substituting the file size expression in~\eqref{filesize:affine} and $\alpha = q^{m-1}$ into~\eqref{Condition}, we obtain
\begin{equation}
k = \Big\lceil q - q(1-\frac{1}{q})^k \Big\rceil,
\end{equation}
which can be rewritten as
\begin{equation} \label{immediate_affine}
k - 1 < q - q(1-\frac{1}{q})^k \leq k.
\end{equation}

For simplicity, we define the following function
$$
F(k) := q(1-\frac{1}{q})^k + k - q,
$$
where $1 \leq k \leq \min \{m, k_0\}$. It can be computed that
\begin{equation}
F(k+1) - F(k) = 1- (1-\frac{1}{q})^k > 0,
\end{equation}
which suggests that $F(k)$ is an increasing function of $k$. Since $F(1)=0$, we have
\begin{equation}
k \geq q - q(1-\frac{1}{q})^k.
\end{equation}

Furthermore, the left-hand side term in~\eqref{immediate_affine} follows from~the condition given in~\eqref{condition_affine}, which completes the proof.
\end{IEEEproof}

For any given prime power $q$, we can find the appropriate~$k_0$ such that Inequality~\eqref{condition_affine} holds. By choosing the parameters $m$ and $\rho$, we can construct FR codes that achieve the Singleton-like bound in~\eqref{SingletonBound}. For example, if $q=16$, we obtain $k_0=6$ in this case. By setting $m=6$ and~$\rho=7$, there exists a~$(112, 16^5, 7)$-FR code attaining the Singleton-like bound in~\eqref{SingletonBound} for $1\leq k \leq 6$. If $q=81$, then it follows that $k_0=13$. We can obtain a~$(1134, 81^{12}, 14)$-FR code that achieves the Singleton-like bound in~\eqref{SingletonBound} for $1\leq k \leq 13$ when $m=13$ and $\rho=14$.

\textit{Remark 3.} One of the key advantages of an affine resolvable design based FR code is that each helper node contributes the same number of packets for node repair, thus achieving load-balancing between helper nodes~\cite{key-22}. Besides the construction based on affine geometries, we can also study affine resolvable designs from mutually orthogonal Latin squares (MOLS)~\cite{key-28}. A \textit{Latin square} of order $n$ is an $n \times n$ array in which each cell contains one point from the set $1,2,\ldots,n$, such that each point occurs exactly once in each row and exactly once in each column. For a Latin square $L$ of order $n$, we denote the $(i, j)$-entry by $L(i, j)$, where $i,j=1,2,\ldots,n$. We say that two Latin squares $L_1$ and $L_2$ of order $n$ are \textit{orthogonal} if, for any $x,y \in \{1,2,\ldots,n\}$, there is a unique cell $(i, j)$ such that $L_1 (i, j) = x$ and $L_2 (i, j) = y$. A set of Latin squares of order $n$, say $L_1, L_2 ,\ldots, L_s$, are said to be \textit{mutually orthogonal} if~$L_i$ and $L_j$ are orthogonal for all $1 \leq i < j \leq s$. Suppose that $p$ is a prime and $m$ is a positive integer. In~\cite{key-13}, the authors presented an explicit construction of $p^m-1$ MOLS of order $p^m$, which can be used to design a $(\rho p^m, p^m, \rho\leq p^m-1)$-FR code $\cC$ with file size
\begin{equation}
M_k(\cC) = kp^m-\binom{k}{2},
\end{equation}
where $1 \leq k \leq \rho$. Since the expression for supported file size is similar to that in~\eqref{FileSizeSteiner}, we have the following result.

\begin{cor} \label{cor:MOLS}
Let $p$ be a prime and $m$ be a positive integer. There exists a $(\rho p^m, p^m, \rho \leq p^m-1)$-FR code attaining the Singleton-like bound in~\eqref{SingletonBound} for $1\leq k \leq \min\{\rho, \lceil \frac{\sqrt{1+8p^m} - 1}{2} \rceil\}$.
\end{cor}

Consider now $(p,m,\rho) = (5,2,7)$. In this case, we can obtain a $(175,25,7)$-FR code attaining the Singleton-like bound in~\eqref{SingletonBound} for $1\leq k \leq 7$.

\textit{Remark 4.} All the optimal FR codes evaluated in this section satisfy the condition that $k \leq d$, i.e., the repair locality is larger than (or equal to) the reconstruction degree. In this scenario, the Singleton-like bound in~\eqref{SingletonBound} is essentially the same as the upper bound in~\eqref{Papailiopoulos}. Since node repair locality is also an important metric in practical distributed storage systems, we consider in the following section locally repairable FR codes that achieve the upper bound in~\eqref{Papailiopoulos}.

\section{Optimal FR Codes Attaining the Minimum Distance Bound in~\eqref{Papailiopoulos}}

In this section, we investigate FR codes with repair locality $d<k$ that attain the upper bound in~\eqref{Papailiopoulos}, i.e., optimal FR codes that also enjoy the desirable local repair property. Our observation is that for a given FR code $\cC$, the repair locality $d$ can be uniquely determined based on the code structure.%
\footnote{For example, an explicit algorithm for computing the repair locality of FR codes is presented in~\cite{key-29}.}
On the other hand, the reconstruction degree $k$ depends on the size of stored files. As $M_k(\cC)$ is a non-decreasing function of $k$, we can employ $\cC$ to store data objects of certain large sizes such that the corresponding reconstruction degrees are larger than the repair locality. Although this method yields locally repairable FR codes, whether they achieve the minimum distance bound in~\eqref{Papailiopoulos} remains to be studied. Unfortunately, for the code constructions based on resolvable designs in~\cite{key-13}, the authors showed that simply increasing the value of $k$ may result in suboptimal FR codes with respect to the bound in~\eqref{Papailiopoulos}. However, this is not always the case and we present in the subsequent discussion explicit locally repairable FR codes that attain the upper bound in~\eqref{Papailiopoulos}.

Similar to Theorem~\ref{basecondition}, we have the following result.

\begin{lem}
Let $\cC$ be an $(n,\alpha,\rho)$-FR code with repair locality $d$. If the reconstruction degree $k$ satisfies
\begin{equation} \label{Condition2}
k = \Big\lceil \frac{M_k(\cC)}{\alpha} \Big\rceil + \Big\lceil \frac{M_k(\cC)}{d\alpha} \Big\rceil -1,
\end{equation}
then $\cC$ attains the minimum distance bound in~\eqref{Papailiopoulos}.
\end{lem}

In the following, we consider optimal FR codes attaining the bound in~\eqref{Papailiopoulos} from regular graphs with large girth. Moreover, the dual codes can also achieve the bound in~\eqref{Papailiopoulos} for certain scenarios.

\begin{thm} \label{thm:regular2}
Let $\cC$ be an FR code constructed from an $\alpha$-regular graph with girth $g$. Then, $\cC$ attains the upper bound in~\eqref{Papailiopoulos} if the reconstruction degree $k>\alpha$ satisfies the requirements~listed in Table~\ref{k:condition}.
\end{thm}

\begin{IEEEproof}
Recall that $\cC$ is an $(n,\alpha,2)$-FR code with repair locality $d=\alpha$, and its file size is $M_k(\cC) = k\alpha-k+1$ for $1 \leq k\leq g-1$. We will focus on the scenario wherein the reconstruction degree $k$ is strictly larger than $\alpha$, and we consider two different cases depending on the relation between $k$ and $\alpha$.

\textit{Case 1:} $k = a_0\alpha \leq g-1$, where $a_0 > 1$ is a positive integer. In this case, we have
\begin{equation}
\Big\lceil \frac{M_k(\cC)}{\alpha} \Big\rceil + \Big\lceil \frac{M_k(\cC)}{\alpha^2} \Big\rceil = \Big\lceil a_0(\alpha - 1) + \frac{1}{\alpha} \Big\rceil + \Big\lceil a_0 - \frac{a_0\alpha-1}{\alpha^2} \Big\rceil
= a_0\alpha + 1 - \Big\lfloor \frac{a_0\alpha-1}{\alpha^2} \Big\rfloor,
\end{equation}
which equals to $a_0\alpha+1 = k+1$ if
\begin{equation}
0 \leq \frac{a_0\alpha-1}{\alpha^2} < 1.
\end{equation}
It follows that $a_{0} < \frac{\alpha^2 + 1}{\alpha}$. Thus, we obtain $a_{0} \leq \alpha$ since $a_0$ is an integer.

\begin{table*}
\caption{The Requirements on $k$ for A Given Tuple $(\alpha,g)$}
\begin{centering}
\begin{tabular}{|c|c|c|c|}
\hline
Relation among $k$, $\alpha$, and $g$ & \multicolumn{3}{c|}{Requirements}\tabularnewline
\hline
\hline
$k = a_{0}\alpha \leq g-1, a_{0} > 1$ & \multicolumn{3}{c|}{$a_{0} \leq \alpha$}\tabularnewline
\hline
\multirow{2}{*}{$k = a_{1}\alpha + b_{1} \leq g-1, a_{1}\geq 1, 1 \leq b_{1} < \alpha$} & \multirow{3}{*}{$a_{1} < b_{1}$} & $\Delta_{1} \leq 0$ & $\alpha > \frac{b_{1}-1}{b_{1} - a_{1}}$\tabularnewline
\cline{3-4} \cline{4-4}
 &  & \multirow{2}{*}{$\Delta_{1} > 0$} & $\frac{b_{1}-1}{b_{1} - a_{1}} < \alpha \leq \frac{b_{1}-a_{1} - \sqrt{\Delta_{1}}}{2}$
or \tabularnewline
Define $\Delta_{1} := (b_{1}-a_{1})^{2} - 4b_{1} + 4$ &  &  & $\alpha \geq \max \big\{\frac{b_{1}-a_{1} + \sqrt{\Delta_{1}}}{2}, \frac{b_{1}-1}{b_{1}-a_{1}}+1 \big\}$\tabularnewline
\hline
\multirow{2}{*}{$g \leq k = a_{2}\alpha+b_{2}\leq g + \left\lceil \frac{g}{2}\right\rceil - 2, a_{2}\geq 1, 1 \leq b_{2} < \alpha$} & \multirow{3}{*}{$a_{2} < b_{2}$} & $\Delta_{2} \leq 0$ & $\alpha > \frac{b_{2}}{b_{2}-a_{2}}$\tabularnewline
\cline{3-4} \cline{4-4}
 &  & \multirow{2}{*}{$\Delta_{2} > 0$} & $\frac{b_{2}}{b_{2} - a_{2}} < \alpha \leq \frac{b_{2}-a_{2} - \sqrt{\Delta_{2}}}{2}$
or \tabularnewline
Define $\Delta_{2} := (b_{2}-a_{2})^{2} - 4b_{2}$ &  &  & $\alpha \geq \max \big\{\frac{b_{2}-a_{2}+\sqrt{\Delta_{2}}}{2}, \frac{b_{2}}{b_{2}-a_{2}}+1 \big\}$\tabularnewline
\hline
\end{tabular}
\par\end{centering}
\label{k:condition}
\end{table*}

\textit{Case 2:} $k = a_1\alpha + b_1 \leq g-1$, where $a_1 \geq 1$ and $1 \leq b_1 < \alpha$ are positive integers. Similarly, we can compute that
\begin{equation}
\Big\lceil \frac{M_k(\cC)}{\alpha} \Big\rceil = \Big\lceil a_1(\alpha - 1) + b_1 - \frac{b_1 - 1}{\alpha} \Big\rceil = a_1(\alpha - 1) + b_1,
\end{equation}
and
\begin{equation}
\Big\lceil \frac{M_k(\cC)}{\alpha^2} \Big\rceil  = \Big\lceil a_1 + \frac{(b_1 - a_1)\alpha-b_1+1}{\alpha^2} \Big\rceil.
\end{equation}

\smallskip{}

Note that if $b_1 \leq a_1$, then $\lceil \frac{M_k(\cC)}{\alpha^2} \rceil \leq a_1$, and $\lceil \frac{M_k(\cC)}{\alpha} \rceil + \lceil \frac{M_k(\cC)}{\alpha^2} \rceil \leq a_1\alpha + b_1 < k+1$. Therefore, we must have $b_1 > a_1$ in order to generate an FR code attaining the bound in~\eqref{Papailiopoulos}. If $b_1 > a_1$,~we obtain
\begin{equation}
\Big\lceil \frac{M_k(\cC)}{\alpha} \Big\rceil + \Big\lceil \frac{M_k(\cC)}{\alpha^2} \Big\rceil = a_1\alpha + b_1 + \Big\lceil \frac{(b_1 - a_1)\alpha-b_1+1}{\alpha^2} \Big\rceil.
\end{equation}
Then, $\lceil \frac{M_k(\cC)}{\alpha} \rceil + \lceil \frac{M_k(\cC)}{\alpha^2} \rceil = k+1$ holds if
\begin{equation}
0 < \frac{(b_1 - a_1)\alpha-b_1+1}{\alpha^2} \leq 1.
\label{condition:RG1}
\end{equation}

The left-hand side term in~\eqref{condition:RG1} suggests that
\begin{equation}
\alpha > \frac{b_1-1}{b_1-a_1},
\end{equation}
and the right-hand side term can be rewritten as
\begin{equation}
\alpha^2 - (b_1 - a_1)\alpha + b_1 - 1 \geq 0.
\end{equation}

Let $\Delta_{1}:=(b_{1}-a_{1})^{2}-4(b_{1}-1)$. If $\Delta_{1} \leq 0$, the inequality above always holds, and if $\Delta_{1} > 0$, we have
\begin{equation}
\alpha\leq\frac{b_{1}-a_{1}-\sqrt{\Delta_{1}}}{2}  \text{ or } \alpha \geq \frac{b_{1}-a_{1}+\sqrt{\Delta_{1}}}{2}.
\end{equation}

The proof is completed by further taking the supported file size of $\cC$ for $g \leq k \leq g+\lceil\frac{g}{2}\rceil-2$ (i.e., $M_k(\cC) = k\alpha-k$) into consideration.
\end{IEEEproof}

\textit{Remark 5.} In~\cite{key-13,key-15}, the authors considered FR codes designed from regular graphs with given girth that attain the minimum distance bound in~\eqref{Papailiopoulos}, where they focused on either the case $M_k(\cC) = k\alpha-k$ (see e.g. Lemma 15~\cite{key-13}) or $M_k(\cC) = k\alpha-k +1$ (see e.g. Theorem 6~\cite{key-15}). We note that there exist two main distinctions between the codes in~\cite{key-13,key-15} and those evaluated in Theorem~\ref{thm:regular2}. The first distinction is that the reconstruction degree is restricted to $k=g$ in~\cite{key-13} and $k < \frac{g}{2}$ in~\cite{key-15}, yet the optimal FR code in this section is discussed for $1 \leq k \leq g+\lceil\frac{g}{2}\rceil-2$. The second distinction lies in that the relation among the parameters $\alpha$ and $a_i,b_i, 1\leq i \leq 2$ is not investigated in~\cite{key-13,key-15}, which is shown to be important in forming an optimal FR code. Furthermore, their results do not cover the case when $\alpha$ divides $k$. Indeed, both the regular graph based optimal FR codes in~\cite{key-13} and~\cite{key-15} can be viewed as special cases as those in Theorem~\ref{thm:regular2}.


\begin{thm} \label{thm:regular3}
Let $\cC$ be an FR code constructed from an $\alpha$-regular graph with $n$ vertices and girth $g$. Then, the dual code $\cC^t$ attains the upper bound in~\eqref{Papailiopoulos} if one of the following conditions holds
\begin{enumerate}
\item if $\Lambda_1: = \frac{(n+2)(\alpha-2)}{8\alpha-14}$ is an integer such that $0\leq n-4\Lambda_1 \leq g-2$, and the size of stored file is $4\Lambda_1$;
\item if $\Lambda_2: = \frac{n(\alpha-2)+2}{8\alpha-14}$ is an integer such that $1\leq n-4\Lambda_2 \leq g-1$, and the size of stored file is $4\Lambda_2+1$;
\item if $\Lambda_3: = \frac{(n-2)(\alpha-2)}{8\alpha-14}$ is an integer such that $2\leq n-4\Lambda_3 \leq g$, and the size of stored file is $4\Lambda_3+2$;
\item if $\Lambda_4: = \frac{(n-4)(\alpha-2)}{8\alpha-14}$ is an integer such that $3\leq n-4\Lambda_4 \leq g+1$, and the size of stored file is $4\Lambda_4+3$.
\end{enumerate}
\end{thm}

\begin{IEEEproof}
According to Lemma~\ref{dualrelation}, the dual $\cC^t$ is an $(\frac{n\alpha}{2},2,\alpha)$-FR code with file size
\begin{equation}
M_k(\cC^t) = \begin{cases}
n-1, & \text{for } \frac{n\alpha}{2}-2\alpha+1 <k \leq  \frac{n\alpha}{2}-\alpha,\\
n-2, & \text{for } \frac{n\alpha}{2}-3\alpha+2 <k \leq  \frac{n\alpha}{2}-2\alpha+1,\\
\vdots & \vdots \\
n-g+2, & \text{for } (\frac{n}{2}-g+1)\alpha+g-2 <k \leq  (\frac{n}{2}-g+2)\alpha+g-3.
\end{cases}
\end{equation}
In other words, the reconstruction degree of $\cC^t$ is $N_k(\cC)+1$ when decoding a data object of~size $n-k+1, 1\leq k\leq g-1$. Since each storage node in $\cC^t$ has a repair locality $d=2$, we need to prove that
\begin{equation} \label{Condition3}
N_k(\cC) + 1 = \frac{n\alpha}{2} -k\alpha + k = \Big\lceil \frac{n-k+1}{2} \Big\rceil + \Big\lceil \frac{n-k+1}{4} \Big\rceil -1.
\end{equation}

Indeed, we can discuss four different cases depending on the value of $(n-k+1) \textup{ mod 4}$. For example, if $(n-k+1) \textup{ mod 4} \equiv 0$, then we let $n-k+1 = 4\Lambda$, where $\Lambda$ is a positive integer. Since $k=n-4\Lambda+1$, Equation~\eqref{Condition3} can be rewritten as
\begin{equation}
\frac{n\alpha}{2} -(\alpha-1) (n-4\Lambda+1) = 3\Lambda-1,
\end{equation}
which gives that
\begin{equation}
\Lambda = \Lambda_1 = \frac{(n+2)(\alpha-2)}{8\alpha-14}.
\end{equation}

Therefore, if $\Lambda_1$ is an integer such that $0\leq n-4\Lambda_1 \leq g-2$, then $\cC^t$ attains the upper bound in~\eqref{Papailiopoulos} when storing a data object of size $4\Lambda_1$. In addition, the other three cases can be obtained following a similar procedure, which completes the proof.
\end{IEEEproof}

\smallskip{}

For example, we consider a \textit{cycle graph} with $n \geq 5$ vertices. There exists an $(n,2,2)$-FR code that attains the upper bound in~\eqref{Papailiopoulos} when the size of stored file is $M=5$. If $(n,\alpha,g) = (18,3,12)$, then there exists a $(27,2,3)$-FR code that achieves the upper bound in~\eqref{Papailiopoulos} for $M=8$ and $M=9$, respectively.

\begin{cor} \label{cor:regular2}
Let $\cC$ be an FR code derived from an $\alpha$-regular graph with $n$ vertices and girth $g$. Then, $\cC^t$ attains the upper bound in~\eqref{Papailiopoulos} if one of the followings holds
\begin{enumerate}
\item if $\Lambda'_1: = \frac{n(\alpha-2)+2(\alpha-3)}{8\alpha-14}$ is an integer such that $g-1\leq n-4\Lambda'_1 \leq g+\lceil\frac{g}{2}\rceil-3$, and the size of stored file is $4\Lambda'_1$;
\item if $\Lambda'_2: = \frac{n(\alpha-2)}{8\alpha-14}$ is an integer such that $g\leq n-4\Lambda'_2 \leq g+\lceil\frac{g}{2}\rceil-2$, and the size of stored file is $4\Lambda'_2+1$;
\item if $\Lambda'_3: = \frac{n(\alpha-2)-2(\alpha-1)}{8\alpha-14}$ is an integer such that $g+1\leq n-4\Lambda'_3 \leq g+\lceil\frac{g}{2}\rceil-1$, and the size of stored file is $4\Lambda'_3+2$;
\item if $\Lambda'_4: = \frac{n(\alpha-2)-2(\alpha-3)}{8\alpha-14}$ is an integer such that $g+2\leq n-4\Lambda'_4 \leq g+\lceil\frac{g}{2}\rceil$, and the size of stored file is $4\Lambda'_4+3$.
\end{enumerate}
\end{cor}

\begin{IEEEproof}
The proof follows similar steps as that given in Theorem~\ref{thm:regular3}, where the only difference is that the supported file size of $\cC$ is $M_k(\cC) = k\alpha-k$ for $g \leq k \leq g+\lceil\frac{g}{2}\rceil-2$.
\end{IEEEproof}

\textit{Remark 6.} In~\cite{key-30}, the construction of graphs with an arbitrary large girth and of large size is proposed, which in conjunction with the results above yields a large family of distance optimal locally repairable FR codes.

\section{An Improved Upper Bound on the Minimum Distance of FR Codes}

In this section, we provide a new upper bound on the minimum distance of FR codes, which is shown to be tighter than the existing bounds in~\eqref{SingletonBound} and~\eqref{Papailiopoulos}, especially when the size of stored file is sufficiently large. We note that the upper bounds in~\eqref{SingletonBound} and~\eqref{Papailiopoulos} hold for all erasure codes, yet FR codes have a more stringent requirement that the repair process is completed by pure data transfer. Moreover, both the two upper bounds have no connection with the repetition degree $\rho$, which is an important metric in defining an $(n,\alpha,\rho)$-FR code.

\subsection{An Improved Upper Bound}

We start by introducing the following useful lemma, which provides another upper bound on the supported file size of FR codes.

\begin{lem} (\cite{key-16}) \label{lem:dualboundMk}
Let $\cC$ be an $(n,\alpha,\rho)$-FR code with $\theta=\frac{n\alpha}{\rho}$. Define the function $\psi(\ell)$~recursively as
\begin{equation} \label{eq:Dualbound}
\psi(1) := \rho, \ \psi(\ell+1) := \psi(\ell) + \rho - \Big\lceil \frac{\alpha \psi(\ell) - \ell \rho}{\theta-\ell}\Big\rceil,
\end{equation}
for $\ell=1,2,\ldots, \theta-1$. Then, the supported file size of $\cC^t$ is upper bounded by
\begin{equation}\label{eq:Dualbound1}
M_\ell(\cC^t) \leq \psi(\ell),
\end{equation}
and for $k = 1,2,\ldots, n$, the supported file size of $\cC$ is upper bounded by
\begin{equation} \label{eq:Dualbound2}
M_k(\cC) \leq \sum_{\ell=1}^\theta \mathbb{I}( k > n - \psi(\ell)),
\end{equation}
where $\mathbb{I}(\cdot)$ is the indicator function equal to $1$ if the condition is true and $0$ otherwise.
\end{lem}

It is shown in~\cite{key-16} that the upper bound in~\eqref{eq:Dualbound2} can be tighter than the original bound in~\eqref{eq:FRbound} for certain code parameters. Thus, a safe conclusion can be obtained by taking the~minimum of the two bounds.

We state the main result of this section in the following theorem.

\begin{thm}
Let $\cC$ be an $(n,\alpha,\rho)$-FR code with $\theta=\frac{n\alpha}{\rho}$. When storing a data object of size $M \leq \theta$, the minimum distance $d_{\min}$ of the system based on $\cC$ is upper bounded by
\begin{equation} \label{eq:newbound}
d_{\min} \leq \min \Big( \psi(\theta - M + 1), \sum_{k=1}^n \mathbb{I}(\varphi(k) > M - 1) \Big),
\end{equation}
where the functions $\varphi(\cdot)$ and $\psi(\cdot)$ are defined in~\eqref{eq:FRbound2} and~\eqref{eq:Dualbound}, respectively.
\end{thm}

\begin{IEEEproof}
From the dual perspective, we observe that any $\theta - M + 1$ packets in $\cC$ are distributed across at least $M_{\theta - M + 1}(\cC^t)$ storage nodes, and if these nodes fail simultaneously, we can only obtain $M-1$ distinct coded packets from the surviving nodes. On the other hand, if the number of node failures is strictly less than $M_{\theta - M + 1}(\cC^t)$, then we will obtain at least $M$ distinct packets. By definition, the system minimum distance is
\begin{equation}
d_{\min} = M_{\theta - M + 1}(\cC^t),
\end{equation}
when the size of stored file is $M$.

According to Lemma~\ref{lem:dualboundMk}, we have $M_{\theta - M + 1}(\cC^t) \leq \psi(\theta - M + 1)$ on one hand, and on the~other hand,
\begin{equation}
M_{\theta - M + 1}(\cC^t) \leq \sum_{k=1}^n \mathbb{I}(\theta - M + 1 > \theta - \varphi(k)) = \sum_{k=1}^n \mathbb{I}(\varphi(k) > M - 1),
\end{equation}
which completes the proof.
\end{IEEEproof}

For example, we consider now an $(8,3,2)$-FR code, which can be constructed from a $3$-regular graph on $8$ vertices. It can be verified that
\begin{gather*}
\varphi(1)=3,\  \varphi(2)=5,\ \varphi(3)=7,\ \varphi(4)=9,\\
\varphi(5)=10, \varphi(7)=11,\ \varphi(7)=\varphi(8)=12.
\end{gather*}
Moreover, we can compute that
\begin{gather*}
\psi(1)=2,\  \psi(2)=3,\ \psi(3)=4,\ \psi(4)=5,\ \psi(5)=\psi(6)= 6,\\
\psi(7)=\psi(8)=\psi(9)=7,\ \psi(10)=\psi(11)=\psi(12)=8.
\end{gather*}
For $M=5,6\ldots,11$, the minimum distance of the corresponding system is upper bounded by
$$
d_{\min} \leq \begin{cases}
\min  \big( \psi(8), \sum_{k=1}^8 \mathbb{I}(\varphi(k)>4) \big) = 7, & \text{for } M=5,\\
\min  \big( \psi(7), \sum_{k=1}^8 \mathbb{I}(\varphi(k)>5) \big) = 6, & \text{for } M=6,\\
\min  \big( \psi(6), \sum_{k=1}^8 \mathbb{I}(\varphi(k)>6) \big) = 6, & \text{for } M=7,\\
\min  \big( \psi(5), \sum_{k=1}^8 \mathbb{I}(\varphi(k)>7) \big) = 5, & \text{for } M=8,\\
\min  \big( \psi(4), \sum_{k=1}^8 \mathbb{I}(\varphi(k)>8) \big) = 5, & \text{for } M=9,\\
\min  \big( \psi(3), \sum_{k=1}^8 \mathbb{I}(\varphi(k)>9) \big) = 4, & \text{for } M=10,\\
\min  \big( \psi(2), \sum_{k=1}^8 \mathbb{I}(\varphi(k)>10) \big) = 3, & \text{for } M=11.
\end{cases}
$$

\smallskip{}

\begin{table*}
\protect\caption{Optimal FR Codes Attaining the Proposed Upper Bound in~\eqref{eq:newbound}}
\begin{centering}
\begin{tabular}{|c|c|c|}
\hline
Method & Code parameter & File size \tabularnewline
\hline
\hline
$(n, r)$-Turán graph & $(\frac{n^2(r-1)}{2r},2,\frac{n(r-1)}{r})$  & $\frac{n}{r}+1 \leq M \leq n$ \tabularnewline
\hline
Regular graph with large girth & $(\frac{n\alpha}{2},2,\alpha)$ & $n-\alpha+1 \leq M \leq n$ \tabularnewline
\hline
Generalized quadrangle & $\big( (t+1)(st+1),(s+1),(t+1) \big)$ & $s(st+t+1)-1 \leq M \leq (s+1)(st+1)$ \tabularnewline
\hline
Transversal design & $(\alpha^2, \rho, \alpha)$ & $(\rho-1)\alpha+1 \leq M \leq \rho\alpha$ \tabularnewline
\hline
\end{tabular}
\par\end{centering}
\label{Summary:OFRC3}
\end{table*}

\smallskip{}

\textit{Remark 7.} In~\cite{key-18}, several families of FR codes attaining the bound on the supported file size in~\eqref{eq:FRbound} are presented for certain reconstruction degrees. In this case, we emphasize that the duals of these FR codes attain our proposed upper bound on the minimum distance in~\eqref{eq:newbound}, which are listed in Table~\ref{Summary:OFRC3}.%
\footnote{We refer interested readers to~\cite{key-18} for more details about the definitions of related designs and the requirements on the code parameters.}
Furthermore, it is worth noting that these FR codes are optimal with respect to the bound in~\eqref{eq:newbound} for some large file sizes, i.e., the system storage efficiency is considerable.

\begin{figure}
\centering{}\includegraphics[scale=0.765]{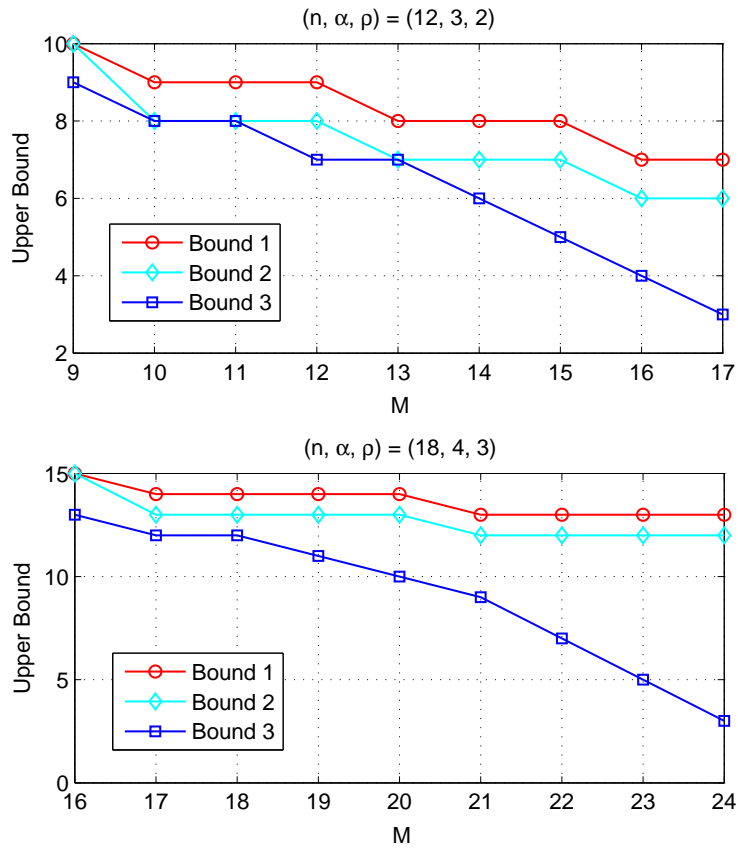}
\caption{Comparison of our proposed bound and the existing bounds, where bound 1, bound 2 and bound 3 refer to the upper bounds in~\eqref{SingletonBound},~\eqref{Papailiopoulos} and~\eqref{eq:newbound}, respectively.}
\label{BoundComparison}
\end{figure}

\subsection{Comparison with Existing Bounds}

In this subsection, we compare the proposed upper bound in~\eqref{eq:newbound} on the minimum distance~of FR codes with the existing bounds in~\eqref{SingletonBound} and~\eqref{Papailiopoulos}. Fig.~\ref{BoundComparison} depicts the three upper bounds of two explicit FR codes with different file sizes. We can observe that our new bound is tighter than the existing bounds in general and the gap becomes wider as the file size $M$ increases. The reason for this gap is that while the upper bounds in~\eqref{SingletonBound} and \eqref{Papailiopoulos} become  looser for larger values of $M$, the recursive bounds in~\eqref{eq:Dualbound1} and~\eqref{eq:Dualbound2} tend to be tighter for smaller reconstruction degrees.

\textit{Remark 8.} In~\cite{key-13,key-15}, the authors considered upper bounds on the minimum distance of a special class of FR codes wherein each storage node belongs to a local FR code that operates~on a smaller symbol set. Specifically, let $\cC$ be an $(n,\alpha,\rho)$-FR code with each node being a part of an $(n',\alpha,\rho')$-FR code, where $n \geq n'$ and $\rho \geq \rho'$. Then, the minimum distance $d_{\min}$ of the~system based on $\cC$ is upper bounded by
\begin{equation} \label{eq:Subound}
d_{\min} \leq n - \Big\lceil n'(1-\frac{1}{\rho'}) \big\lfloor \frac{(M-1)\rho'}{n'\alpha} \big\rfloor + \frac{M}{\alpha} \Big\rceil + 1,
\end{equation}
when the size of stored file is $M$. However, the upper bound in~\eqref{eq:Subound} applies only for FR codes with specific local structures. We show in this section that the minimum distance of an FR code is essentially equivalent to the supported file size of the corresponding dual FR code, and thus our proposed bound holds for any FR code with all possible file sizes. Moreover, the proposed new bound in~\eqref{eq:newbound} can be tighter than the upper bound in~\eqref{eq:Subound} for certain parameter regimes.%
\footnote{Note that there may exist scenarios where the upper bound in~\eqref{eq:Subound} is tighter than the proposed bound in~\eqref{eq:newbound}, implying that the upper bounds on the supported file size in~\eqref{eq:Dualbound1} and~\eqref{eq:Dualbound2} can be improved in such cases.}
As a concrete example, consider an $(18,3,3)$-FR code that obtained from the disjoint union of two copies of a $(9,3,3)$-FR code. For $M=17$, the upper bound in~\eqref{eq:Subound} suggests that the system minimum distance satisfies $d_{\min} \leq 7$, while our proposed bound gives $d_{\min} \leq 5$. Indeed, this bound can be achieved by the following FR code with blocks
\begin{gather*}
\{1,4,7\},\  \{2,5,9\},\ \{3,6,8\},\ \{10,13,16\},\  \{11,14,18\},\ \{12,15,17\},\\
\{1,6,9\},\  \{2,4,8\},\ \{3,5,7\},\ \{10,15,18\},\  \{11,13,17\},\ \{12,14,16\},\\
\{1,5,8\},\  \{2,6,7\},\ \{3,4,9\},\ \{10,14,17\},\  \{11,15,16\},\ \{12,13,18\}.
\end{gather*}
It can be verified that any $14$ storage nodes contain at least $17$ distinct coded packets, and there exist $13$ nodes containing only $16$ distinct packets. Therefore, the minimum distance for $M=17$ is $d_{\min} = 5$.

\section{Conclusion}

In this paper, we consider upper bounds on the minimum distance of FR codes. We present a list of optimal FR codes that attain the Singleton-like bounds in~\eqref{SingletonBound} and~\eqref{Papailiopoulos} respectively, which extend the existing constructions to larger sets. Moreover, we develop an improved upper bound on the minimum distance, which is derived from the dual perspective and holds for all possible FR codes. We show that the proposed bound is tighter than the existing upper bounds in general, and present several optimal FR codes attaining the improved bound.

It is worth mentioning that the upper bound on the minimum distance is essentially equivalent to the upper bound on the supported file size of FR codes. Although some upper bounds have been presented in the literature, a tighter upper bound on the file size (resp. minimum distance) is an interesting direction for future studies. It would also be interesting to construct explicit FR codes that achieve the new bounds.

\end{document}